\documentclass[aip, amsmath,amssymb,
 reprint,%
]{revtex4-1}
\usepackage{mathptmx}
\usepackage{graphicx}
\usepackage{dcolumn}
\usepackage{bm}
\usepackage[utf8]{inputenc}
\usepackage{mdframed}
\usepackage{epsfig}
\usepackage{graphicx}
\usepackage{hyperref}
\usepackage{boxedminipage}
\hypersetup{colorlinks=true,	citecolor=blue,	linktocpage=true}
\usepackage{palatino}
\usepackage{euler,latexsym,amsmath}
\usepackage{rotating}
\usepackage{xspace}
\usepackage{color}
\newcommand{\dfdx}[2]{\left(\frac{\partial #1}{\partial #2}\right)}
\newcommand{\bey}[1]{\begin{eqnarray} \label{#1}}
\newcommand{\eey}{\end{eqnarray}}
\newcommand{\beyn}{\begin{eqnarray*}}
\newcommand{\eeyn}{\end{eqnarray*}}
\newcommand{\beq}[1]{\begin{equation} \label{#1}}
\newcommand{\eeq}{\end{equation}}
\newcommand{\bet}{\begin{table}[tb]\centering}
\newcommand{\ent}{\end{table}}
\newcommand{\bef}{\begin{figure}[tb]\centering}
\newcommand{\enf}{\end{figure}}

\newcommand{\dd}{{\mathrm d}}

\newcommand{\eg}{{e.g.},\xspace}
\newcommand{\ie}{{i.e.},\xspace}

\newcommand{\etal}{{\itshape et al.}\xspace}

\usepackage[colorinlistoftodos, textwidth=20mm]{todonotes}
\makeatletter
\def\@email#1#2{%
 \endgroup
 \patchcmd{\titleblock@produce}
  {\frontmatter@RRAPformat}
  {\frontmatter@RRAPformat{\produce@RRAP{*#1\href{mailto:#2}{#2}}}\frontmatter@RRAPformat}
  {}{}
}%
\makeatother
\begin{document}
\title[Critical crystal nuclei]{The Second Gibbs Paradox}
\author{Daan Frenkel}
 \affiliation{Yusuf Hamied Department of Chemistry, University of Cambridge, Lensfield Road, Cambridge CB2 1EW, United Kingdom. 
 \texttt{df246@cam.ac.uk}
}
\date{\today}
\begin{abstract}
Gibbs's monumental article on the Equilibrium of Heterogeneous Substances  contains  paradoxical sentence stating that, for  a crystallite in equilibrium with a fluid, the chemical potential of the solid will not be equal to that of the fluid  if the surface free-energy density differs from the mechanical surface-tension.
How can this be? 
After all, in chemical equilibrium the chemical potential of any species should be the same throughout the system. 

This ``Second Gibbs Paradox'' has intrigued many authors.
In the present paper I sketch my interpretation of the approach of Gibbs, and that of Mullins (J. Chem. Phys, 81, 1436–1442, 1984), which accounts for the possibility of vacancies and interstitials.
I argue that a consistent treatment of point defects in a critical nucleus is essential for clarifying the meaning of the chemical potential of the nucleus. 
My paper lacks the rigor of Gibbs or Mullins, but will hopefully  be more accessible for scientists who think primarily in terms of atoms and molecules.
In my attempt, I am motivated by a quote that is sometimes attributed to Paul Val\'{e}ry:
{\em ``The glass must be absolutely transparent for one to perceive the mud at the bottom.''} 
\end{abstract}
\maketitle

\section{Preamble}
The style of writing of J. Willard Gibbs was concise. 
So much so that in 1892 Rayleigh wrote to Gibbs, asking him to write an extended version of his famous paper on the Equilibrium of Heterogeneous Substances~\cite{gibbs1928}, because {\em ``...[it] is too condensed and too difficult for most, I might say all, readers.'' }~\cite{wilson1945}.
Gibbs never wrote the didactical paper that Rayleigh wanted: he had other things on his mind -- something he called "Statistical Mechanics" (published ten years later~\cite{gibbs1902}) of which he said: {\em I do not know that I shall have anything particularly new in substance... }
As Gibbs never returned to his work of the 1870's, some of the more striking things that Gibbs wrote about the equilibrium between crystals and fluids have  attracted less attention than they deserve.
In the present article, I discuss a  topic that Gibbs touched upon:  the equilibrium between a small crystal and a fluid (we would now call a small crystal in unstable equilibrium with a fluid a {\em critical nucleus}).
I will sketch both the approach of Gibbs, and that of Mullins~\cite{mullins1984}, which accounts for the possibility of vacancies and interstitials, and I hope to resolve the paradox. 
\section{The Chemical Potential}
The introduction of the concept of the chemical potential ranks among the most important contributions of Gibbs to science. 
Gibbs called this quantity simply the {\em potential} or, in cases where there could be confusion with the electrical of gravitational potential, the {\em intrinsic potential} of a species (see ref.~\onlinecite{gibbs1928} -- paragraph below Eqn. 234).

The term {\em chemical potential} was apparently coined by the rather opinionated (co-)founder  of the Journal of Physical Chemistry (1896), Wilder D Bancroft (see refs.~\onlinecite{baierlein2001,NAS_BiographicalMemoirs}).
Bancroft was a student of Ostwald and van 't Hoff, and the Journal of Physical Chemistry was probably inspired by the Zeitschrift f\"{u}r Physikalische Chemie (1887) created by Ostwald, van 't Hoff and Arrhenius.
Bancroft corresponded with Gibbs, and he had high hopes that the {\em chemical} potential would have a transformative impact on {\em all} of Chemistry:
{\em ``So far as I can see at present, our only hope of converting organic chemistry, for instance, into a rational science lies in the development and application of the idea of the chemical potential.''} (quoted in ref.~\onlinecite{baierlein2001}).
It is my personal opinion [DF] that the use of the adjective ``chemical'' for the intrinsic potential was at best moderately successful in exciting synthetic organic chemists, yet it limited the popularity of the concept among some of the more parochial students in Physics.
\section{Introduction}
Gibbs was well aware of the fact that the correct description of the mechano-chemical equilibrium between a crystalline solid and a fluid containing the same component requires much care~\cite{gibbs1928}, much more, it seemed, than the description of equilibrium between two coexisting fluid phases.
However, in his writings on the subject, Gibbs left out any sentence or even word that he considered redundant. 
The result is that some passages in Gibbs's "Equilibrium of Heterogeneous Substances" are not very transparent to mere mortals. 
This is true in particular for Gibbs's description of the equilibrium between crystals and fluids.
A consequence of Gibbs extreme conciseness is that it took a long time before the implications of some of his findings were spelled out by other authors.
In this context, a key role was played by the continuum mechanics  community, including  authors as such Cahn and Larch\'{e}~\cite{cahn1980,larche1973, larche1978,larche1982}, and Mullins~\cite{mullins1984}. 
More recently, the consequences of Gibbs's insights for the simulation of crystal nucleation were considered, but not fully resolved~\cite{cacciuto2005stresses,montero2020young, montero2022thermodynamics}. 

The present paper contains one result that is ``new'', in the sense that, in the literature,  I have not been able to find  my ``particle-based'' discussion of the chemical potential of a crystal nucleus.
Most other results described below can be found in the literature (see \eg~\onlinecite{dipasquale2025}). 
My argument is hopefully helpful for  people  who, like me,  believe that all material properties should follow from the atomic hypothesis, rather than from a continuum-mechanics description.
Larch\'{e} and Cahn used such a continuum-mechanics approach to explain why there is no problem with the chemical potential of a crystal nucleus~\cite{larche1973,larche1978,larche1982}.
If ``understanding'' means ``being able to explain a physical phenomenon at a birthday party'' (not recommended), then I never fully understood the theory of refs.~\cite{larche1973,larche1978,larche1982}.

To set the problem, I go back to the work of Gibbs~\cite{gibbs1928} and Mullins~\cite{mullins1984} in order to clarify why the correct description of the coexistence between a fluid and a small crystallite is such a minefield. 
Part of the problem for modern readers is that Gibbs tried to avoid the concept of a chemical potential for the individual components inside a solid, as he assumed that there was no mass transport in a solid, writing:
{\em "...it will be more in accordance with our method hitherto, if we consider the solid to have only a single independently variable component of which the nature is represented by the solid itself."} 
In modern language: the composition of a multi-component solid, once formed, is fixed. 
It cannot respond to the change in temperature or chemical potential, except by depositing new solid material with the correct composition,  replacing the original solid, which is dissolved.
Gibbs makes this explicit as follows:
{\em "...the temperature and pressure and the ([DF] chemical) potentials for all the actual components of the solid must have a constant value in the solid at the surface where it meets the fluid. }
This sentence is prescient, as is usual with Gibbs.
Its meaning will hopefully become more clear  at the end of section~\ref{ssec:chempot}.

Yet, since the discovery of vacancies and interstitials in equilibrium solids, we know that mass transport in solids {\em is} possible (although not necessarily simple~\cite{frechette2021}), and hence there should be no problem defining the chemical potential of a species inside a solid -- be it that Gibbs's point still holds that regrowth rather than diffusion is usually the fastest way to change the composition of a solid.
Yet, as we shall see in Sec.~\ref{ssec:chempot}, defining the chemical potential inside a crystal nucleus requires the existence of point defects. 
If Gibbs had been aware of the possibility of point defects in equilibrium crystals, he would have resolved the chemical potential paradox that I will discuss in section~\ref{ssec:virtual}  150 years ago.

\section{The critical crystal nucleus}
The following discussion is based on the original derivation of Gibbs, leading to his Eqn. 661 in ref.~\onlinecite{gibbs1928} .
I am well aware that  ``simplifying'' what Gibbs wrote, almost inevitably leads to errors: these then are mine, not those of Gibbs.
There is also a real difference: Gibbs assumed that the effect of the difference between surface free-energy density and surface stress (see below), can be ignored in practice. 
I do not make this assumption, because there are cases where the difference is important.

First a disclaimer: to my knowledge, Gibbs never used the word {\em crystal nucleus}, nor did he mention the process of nucleation. 
However, in the theoretical framework that he presents in the section on ``Surface of Discontinuity between Solids and Fluids'' of ref.~\onlinecite{gibbs1928}, he describes the very object that we would now call the critical crystal nucleus. 
\subsection{Accretion/dissolution and deformation}
\begin{figure}[htb]
\centering
\includegraphics[width=\columnwidth]{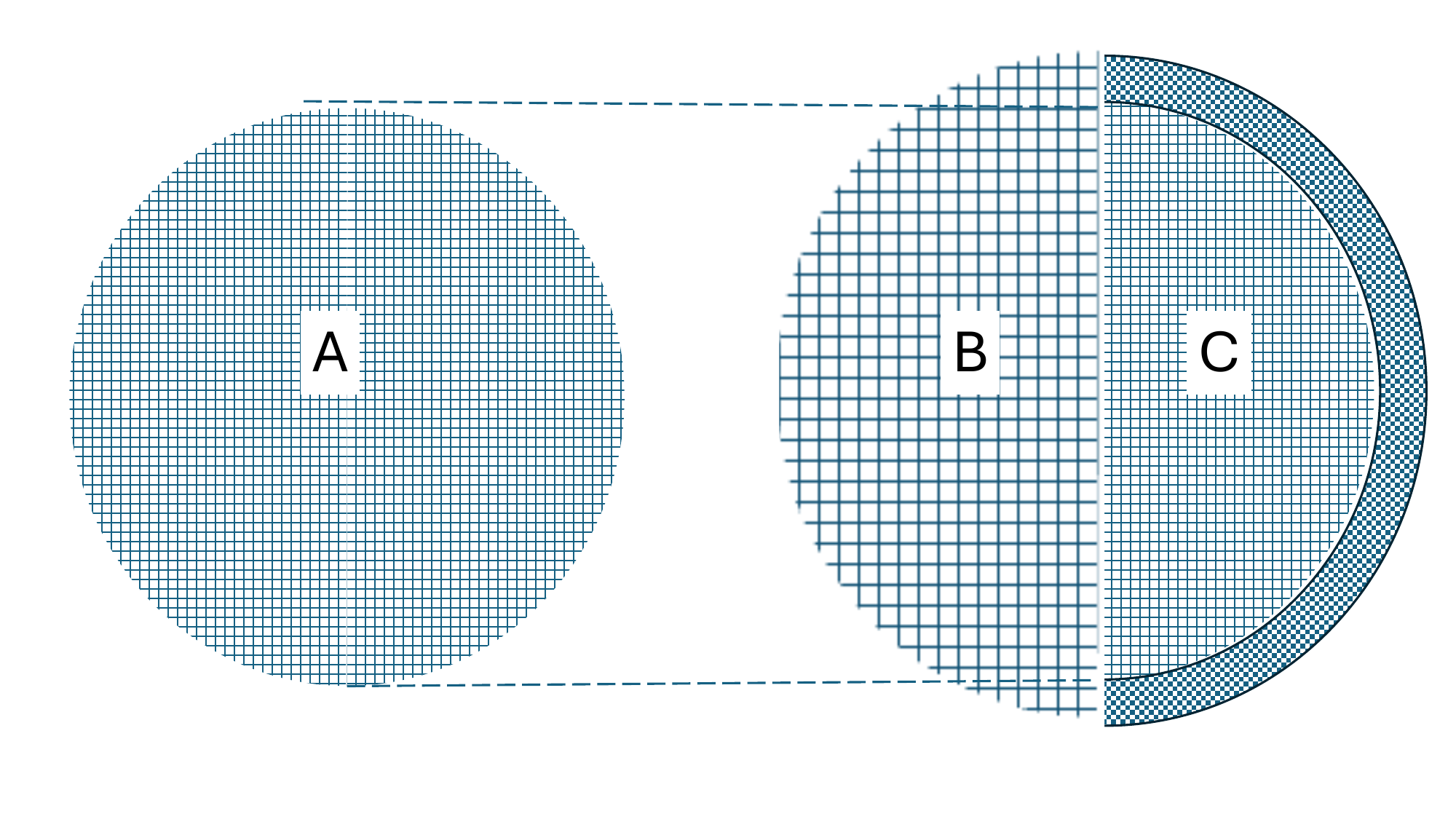}
\caption{A crystallite (A) can expand either through expansion (B) or through accretion (C).
In the case of expansion, the number of lattice sites remains the same, but the lattice spacing changes. 
In the case of accretion, new lattice sites are added.}
\label{fig:accretion_expansion}
\end{figure}
When a small crystallite grows due to accretion, the curvature of its surface changes, but otherwise, its physical properties remain unchanged -- apart from small curvature corrections that I will briefly mention later.
In contrast (see Figure~\ref{fig:accretion_expansion}), if a crystallite expands or contracts at a constant number of lattice sites, the structure of its surface is deformed and work has to be performed against the surface stress $t$ (see below). 
We denote the surface free-energy density by $\sigma $. 
For fluids, $\sigma = t$, because a fluid droplet can create/remove surface when it is deformed (for instance, when a spherical droplet is made ellipsoidal). 
In contrast, if a solid is deformed without changing the arrangements of its lattice points,  it will strain its surface, but cannot renew it. 
The importance of treating the number of lattice sites as an independent thermodynamic variable is discussed refs.~\onlinecite{mladek2007,mladek2008,purohitKofke2018,sprik2025}
\subsection{Virtual volume change}\label{ssec:virtual}
Let us first focus on an infinitesimal change $\Delta V$ in the size of a spherical crystal due to the addition (or removal) of matter from the fluid (\ie accretion or dissolution).
The case of a faceted crystal is more complicated, but not conceptually different. 
We assume for convenience that $\Delta V$  is positive. 
Then the surface area $A$ of the crystallite increases with $\Delta V$.
For a spherical crystal with radius $r$, we have:
\begin{equation}
\Delta V = 4\pi r^2 dr ,
\end{equation}
and
\begin{equation}
\Delta A = 8 \pi r dr =\frac{2\Delta V}{r}.
\end{equation}
The change in surface free energy is then
\begin{equation}\label{eq:VvsA}
\Delta F_{\rm surface} =  \frac{2\sigma }{r}\Delta V .
\end{equation}
Actually, the situation is more subtle than that, as $\sigma $ depends on the radius of curvature: 
\begin{equation}\label{eq:curved-surface}
\sigma (r)\approx\sigma _\infty +\frac{C_1}{r} +\frac{C_2}{r^2} \;,
\end{equation}
where $\sigma _\infty$ is the surface free-energy-density of a flat surface. 
Eqn.~\ref{eq:curved-surface} implicitly contains all terms of the Helfrich Hamiltonian~\cite{helfrich1973}, that is: the radius of spontaneous curvature of the surface, its bending stiffness, and its Gaussian stiffness (see ref.~\onlinecite{blokhuis1992}).

Then
\begin{equation}
F_{\rm surface} = 4\pi r^2 \sigma(r) = 4\pi r^2 \sigma _\infty +4\pi r C_1  +4\pi C_2,
\end{equation}
and the change of $F_s$ due to the addition  (``accretion'') or removal ("dissolution") of a solid volume $\Delta V$ is:
\begin{equation}
\Delta F_s= \left(\frac{2\sigma _\infty}{r} + \frac{C_1}{r^2}\right)\Delta V .
\end{equation}
Note, that this free energy variation is different from what we get if we change the volume of the solid at constant number of particles, \ie by elastic deformation.
In that case, we stretch/compress not only the bulk, but also the surface, and hence we have to account for the change of $\sigma _\infty,C_1$ and $C_2$ with this surface distortion. 
In what follows, I shall mostly ignore the curvature dependence of $\sigma $, not because it is irrelevant, but because it complicates, yet does not change, the main conclusions below. 
With this simplification,  consider what happens to the total Helmholtz free energy of the system if we change the volume of the crystal by {\em accretion}. 
Let us first consider the contribution of the solid. 
We denote the bulk free-energy density of the solid by $f_{\rm solid}$; it is a function of the crystal density and the temperature.  
Then the free-energy change of the crystallite, including the change in surface free energy due to accretion is:
\begin{equation}
\Delta F_{\rm solid} = \left[ f_{\rm solid} +  \frac{2\sigma }{r}\right] \Delta V_{\rm solid} .
\end{equation}
From now on, I focus on the one-component case.
Next consider the fluid, which behaves like a large reservoir. 
Its free energy changes because of the volume change of the solid, and because the growing solid removes particles from the fluid. Then
\begin{eqnarray}\label{eq:DF_solid}
\Delta F_{\rm  fluid} &=&  \mu_{\rm fluid} \Delta N_{\rm fluid} - P_{\rm fluid}\Delta V_{\rm fluid}\nonumber\\
 &=&  \mu_{\rm fluid} \Delta N_{\rm fluid} +P_{\rm fluid}\Delta V_{\rm solid} ,
\end{eqnarray}
where, in the second line I have used the fact that $\Delta V_{\rm fluid}=- \Delta V_{\rm solid}$.
However, 
\begin{equation}\label{eq:DeltaN}
\Delta N_{\rm fluid} = -\Delta N_{\rm solid} = -\rho_{\rm solid} \Delta V_{\rm solid} ,
\end{equation}
where $\rho_{\rm solid}$ is the number density of the solid.
Note that the condition $\Delta N_{\rm fluid} = -\Delta N_{\rm solid} $ implies that Eqn.~\ref{eq:DeltaN} only holds for a nucleus bounded by the Gibbs {\em equimolar} surface.
Note that if Gibbs would have chosen another dividing surface with a non-zero superficial density $\Gamma$, then the relation between $\Delta N_{\rm fluid}$ and $\Delta N_{\rm solid}$ would have been of the form 
\begin{equation}
\Delta N_{\rm fluid} = -\Delta N_{\rm solid} -(2/r)\Gamma \Delta V_{\rm solid}.
\end{equation}
Clearly, in the present case the equimolar dividing surface is simpler, certainly for a one-component system.
Using Eqn.~\ref{eq:DeltaN}, we then have:
\begin{eqnarray}
\Delta F_{\rm  fluid} &=& - \mu_{\rm fluid} \rho_{\rm solid} \Delta V_{\rm solid} +P_{\rm fluid}\Delta V_{\rm solid}\nonumber\\
 &=& \left[- \mu_{\rm fluid} \rho_{\rm solid}  +P_{\rm fluid}\right]\Delta V_{\rm solid}.
\end{eqnarray}
Now we add the changes of the solid and fluid free energies:
\begin{equation}\label{eq:Delta-F}
\Delta F_{\rm total} =\left[ f_{\rm solid} +  \frac{2\sigma }{r}  -\mu_{\rm fluid} \rho_{\rm solid}  +P_{\rm fluid}\right]\Delta V_{\rm solid}.
\end{equation}
Note how Gibbs has cunningly avoided introducing the chemical potential of the solid. 
When the crystallite is in  equilibrium with the fluid (possibly unstable equilibrium, as in the  case of a critical nucleus), we must have $\Delta F_{\rm total}=0$, and hence,
\begin{equation}\label{eq:mu-liq}
\mu_{\rm fluid}=  \frac{f_{\rm solid} +P_{\rm fluid}+  \frac{2\sigma }{r_c}}{\rho_{\rm solid}},
\end{equation}
where $r_c$ is the radius of the critical nucleus, \ie the nucleus that is in unstable equilibrium with the fluid. 
Eqn.\ref{eq:mu-liq} almost looks familiar: the chemical potential of a fluid in equilibrium with a crystal of radius $r$ is the same as the chemical potential that the solid {\em would} have if it experiences a Laplace pressure equal to $2\sigma /r$ -- except that, as we shall see below, the pressure inside the nucleus is {\em not} equal to  $2\sigma /r$ and hence eqn.~\ref{eq:mu-liq} does not describe the chemical potential of a bulk solid at the temperature and pressure that prevail inside the crystal nucleus.

To see this, let us consider the Gibbs free energy of a bulk solid. It is
\begin{equation}
G^{(B)}=F+P_{\rm solid}V ,
\end{equation}
where we have added the superscript $(B)$ to indicate that this relation holds for a bulk material (but not necessarily for a small cluster).
As  $\dd G^{(B)} = \mu^{(B)}_{\rm solid} \dd N$ at constant $P_{\rm solid}$ and $T$, $G^{(B)} =N\mu^{(B)}_{\rm solid}$. 
\footnote{This is where the difference between bulk and finite systems becomes important: as Hill has argued~\cite{hill1994thermodynamics}, for finite $N$, the Gibbs free energy of a cluster equals $G^{(N)}\equiv$  $N\hat{\mu}$, where $\hat{\mu}$ is not equal to $\mu^{(B)}$. }
Hence
\begin{equation}\label{eq:mu_bulk_s}
\mu^{(B)}_{\rm solid} = \frac{f_{\rm solid} +P_{\rm solid}}{\rho_{\rm solid}}.
\end{equation}
But what is $P_{\rm solid}$? 
It is easiest to use the fact that, in equilibrium, we must have 
\begin{eqnarray}
0&=&\dfdx{F_{\rm solid}+ F_{\rm fluid} +\sigma  A}{V_{\rm solid}}_{N,T} \nonumber\\
&=&P_{\rm fluid}-P_{\rm solid}+\dfdx{ \sigma  A}{V_{\rm solid}}_{N,T}.
\end{eqnarray}
Hence, 
\begin{eqnarray}
P_{\rm solid} &=& P_{\rm fluid}+\dfdx{\sigma  A}{V_{\rm solid}}_{N,T} \\
&=& P_{\rm fluid}+\sigma \dfdx{A}{V_{\rm solid}}_{N,T} + A\dfdx{\sigma }{V_{\rm solid}}_{N,T}\nonumber \\
&=& P_{\rm fluid}+\frac{2\sigma }{r}+ \frac{2A}{r}\dfdx{\sigma }{A}_{N,T}\equiv 
P_{\rm fluid}+ \frac{2t}{r}\nonumber.
\end{eqnarray}
In the last term of this equation, we have defined the surface {\em stress} $t$:
\begin{equation}
t\equiv \sigma + A\dfdx{\sigma }{A}_{N,T}.
\end{equation}
For the solid, we therefore have an extra term in the Laplace pressure:
\begin{equation}
\frac{2A}{r}\dfdx{\sigma }{A}_{N,T} .
\end{equation}
Most crystals are not very compressible, in which case the  effect of $\frac{2A}{r}\dfdx{\sigma }{A}_{N,T}$, the extra term in the Laplace pressure, on the density of the solid is negligible.
However, for very compressible solids (\eg solids close to a solid-solid critical point~\cite{cacciuto2005stresses}), the effect can be large. 
It can even change the sign of the Laplace pressure (see also ~\onlinecite{montero2022thermodynamics}).
For the present discussion of the equilibrium between a crystallite and a fluid,  this extra term has counter-intuitive consequences, as combining Eqns.~\ref{eq:mu-liq} and \ref{eq:mu_bulk_s} seems to imply that the chemical potential of solid and fluid at coexistence are {\em not} equal: 
\begin{eqnarray}\label{eq:mus-ne-mul}
\mu^{\rm (B)}_{\rm solid}&=&  \frac{f_{\rm solid} +P_{\rm fluid}+  \frac{2\sigma }{r} + \frac{2A}{r}\dfdx{\sigma }{A}_{N,T}}{\rho_{\rm solid}}\nonumber\\ 
&=&\mu_{\rm fluid}+ \frac{2(t-\sigma )/r}{\rho_{\rm solid}},
\end{eqnarray}
where, as before,  the superscript $B$ on the chemical potential of the solid indicates that this is the chemical potential that a bulk solid would have if it would be at the same temperature and pressure as the inside of the nucleus. 
Interestingly (but not surprisingly), Gibbs already hints at the result expressed by Eqn.~\ref{eq:mus-ne-mul}, writing (in my ``translation''): {\em if the surface free-energy density differs notably from the true (``mechanical'') surface-tension, the chemical potential of the solid will not be equal to that of the fluid, unless the curvature of the surface vanishes.}~\footnote{This is my ``translation''. 
The sentence that Gibbs actually wrote is (hopefully) equivalent, but requires knowledge of other parts of Gibbs's text: {\em ``if in the case of any amorphous body the value of $\sigma$  differs notably from the true surface-tension , the latter quantity substituted for $\sigma$ in (661) will make the
second member of the equation equal to the true value of $\mu_i^{\rm solid}$, when
the stresses are isotropic, but this will not be equal to the value of $\mu_i^{\rm liquid}$ in case of equilibrium, unless $c_1$ + $c_2$=0. [\ie when the total curvature vanishes]''.} } Actually, Gibbs considers the case of the equilibrium between a fluid and an isotropically-stressed {\em amorphous} solid, presumably a glass. But that opens a whole new can of worms~\cite{vinutha2021}.

The inequality between the chemical potentials of a {\em bulk solid} at the Laplace pressure of the crystal nucleus,  and that of the fluid at coexistence has given rise to much {\em Angst}.
However, there is no real problem with the expression for the nucleation barrier, as the condition for equilibrium has been derived from the Second Law of Thermodynamics, in the form stating that the free energy of the system as a whole is at an extremum (minimum) when solid and fluid are in equilibrium at constant total volume and temperature. 

From Eqns.~\ref{eq:Delta-F} and \ref{eq:mu-liq} we can derive the expression for the barrier of an isotropic (``spherical'') crystallite. 
First, I note that the free energy cost of creating a critical nucleus with volume $V_c=(4\pi/3)r_c^3$ is
\begin{equation}\label{eq:barrier-1}
\Delta F_{\rm critical} =\int_0^{V_c}\left[ f_{\rm solid} +  \frac{2\sigma }{r}  -\mu_{\rm fluid} \rho_{\rm solid}  +P\right]\dd V_{\rm solid}.
\end{equation}
Using Eqn.~\ref{eq:mu-liq}, we can write
\begin{equation}\label{eq:sigma1}
\rho_{\rm solid}\mu_{\rm fluid}- f_{\rm solid} -P = \frac{2\sigma }{r_c},
\end{equation}
and hence
\begin{eqnarray}\label{eq:barrier-2}
\Delta F_{\rm critical} &=&-\frac{2\sigma }{r_c}V_c + 
\int_0^{V_c}  \frac{2\sigma }{r}  \dd V_{\rm solid}\nonumber\\
&=&-\frac{8\pi\sigma }{3}r_c^2 +4\pi\sigma  r_c^2\nonumber\\
&=& \frac{\sigma  A_c}{3} \;,
\end{eqnarray}
where $A_c$ is the equimolar dividing  surface area of the critical nucleus. 
Below, we will derive an expression for the barrier height in terms of the surface of tension. 

The equilibrium condition for a crystal nucleus in contact with a fluid, can also be expressed in terms of the grand-canonical density.
Such a description was used by Mullins~\cite{mullins1984}, and has the advantage that it provides a natural choice for the dividing surface between fluid and crystal. 
Reading Mullins' original derivation is complicated by the fact that it contains typos in some crucial equations.  
However, as the approach is very interesting, I give my own summary below (and I apologize {\em a priori} for my own typos).
\subsection{Mullins route}\label{ssec:mullins}
Unlike the better-known derivations of the nucleation barrier and the critical nucleus size, Mullins derivation~\cite{mullins1984} starts from the observation that in equilibrium, the {\em Grand Potential} $\Omega$ of a system consisting of a (crystal) nucleus and its parent phase must be at an extremum, where the superscript ${\rm tot}$ indicates the tot(al) system. 
\beq{eq:omega}
\Omega^{\rm tot} = E^{\rm tot}-TS^{\rm tot}-\mu_iN_i^{\rm tot} ,
\eeq
where, as before, $E^{\rm tot},S^{\rm tot}$ and $N^{\rm tot}_i$ denote the total energy, total entropy and total number of particles of species $i$, and $\mu_i$ is the imposed chemical potential. 
In Eqn.~\ref{eq:omega}, I have used the Einstein summation convention for repeated indices.
However, in what follows I will focus on the one-component case. 
The condition for (unstable) equilibrium is that any small variation in the size and composition of the nucleus will increase the total grand potential:
\begin{equation}
\left(\delta \Omega^{\rm tot}\right)_{T,V^{\rm tot},\{N_i^{\rm tot}\}} =0 .
\end{equation}
The total volume of the system is equal to the sum of the volume of the crystal nucleus $V^c$ and the remaining (fluid) volume $V^{\rm liq}=V^{\rm tot}-V^c$.
In other words, I assume (again following Gibbs) that the surface of the nucleus occupies no volume -- but it is not the equimolar surface.
However, the surface (denoted by $A$) {\em does} contribute to the total grand potential:
\begin{equation}
\Omega^{\rm tot}=\Omega^{\rm cryst}+\Omega^{\rm liq}+\Omega^{\rm surf} .
\end{equation}
Note that all $\Omega$'s depend on the intensive variables $T$ and $\mu_i$, and on either the volume or the area of the nucleus.
Importantly, using the grand potential makes it immediately clear that the chemical potential is the same throughout the system:
unlike Gibbs, Mullins had no qualms about attributing chemical potentials to the species in a solid.
And they both agreed that the chemical potential of species $i$ adsorbed in the surface is equal to that in the bulk fluid.

In what follows, I assume that the crystal nucleus is spherical. 
Its volume is therefore $V^{\rm cryst}=(4\pi/3) r^3$, where the radius $r$ will be fixed later.

It is convenient to define {\em intensive} grand-canonical energy densities for the crystal and  fluid:
\begin{equation}
\omega^{\rm cryst}\equiv\Omega^{\rm cryst}/V^{\rm cryst} =\frac{\Omega^{\rm cryst}}{(4/3)\pi r^3} .
\end{equation}
and
\begin{equation}
\omega^{\rm liq}\equiv\Omega^{\rm liq}/(V^{\rm tot}-V^{\rm cryst}).
\end{equation}
Importantly, the approach of Mullins does not assume that $\nu_i$,  the number of particles of species $i$ per crystal unit cell, is fixed. 
In other words: the crystal phase may contain vacancies or interstitials. 
We denote the number of unit cells in the crystal by $N_L$. 
The number of particles per unit cell, $\nu_i$, is not fixed. 
Therefore the number of particles in a crystal may vary either due to a change in the number of unit cells, or due to a change in the number of particles per unit cell: 
\beq{eq:dN}
dN^{\rm cryst}_i = N_Ld\nu_i+\nu_idN_L .
\eeq
Similarly, the energy of the crystal may vary either due to a change in the energy per unit cell, or due to a change in the number of unit cells:
\beq{eq:dE}
dE^{\rm cryst} = N_L de+e dN_L,
\eeq
where $e$ denotes the energy per unit cell. 
And finally, the variation of the in the volume of the crystal is the sum of the changes to to the number of unit cells, and in the volume $\mbox{\it v}$ per unit cell. 
Therefore:
\beq{eq:dV}
dV^{\rm cryst} = \mbox{\it v}dN_L + N_Ld\mbox{\it v}
\eeq
This decomposition is crucial for what follows.
We now consider a variation in the grand-canonical potential of the system at constant $\mu_i$ and $T$, making use of the fact that the grand potential of a homogeneous fluid phase satisfies: 
\begin{equation}
d\Omega^{\rm liq}  =  -P^{\rm liq}dV^{\rm liq} -S^{\rm liq}dT + N_i^{\rm liq}d\mu_i  = -P^{\rm liq}dV^{\rm liq}.
\end{equation}
The variation in $\Omega^{\rm cryst}$ is a bit more subtle.
We start from the expression for the variation of, $e$, the energy per unit cell of a system, as a function of the entropy, volume and number of particles per unit cell ( $s$, $\mbox{\it v}$ and $\nu_i$):
\begin{equation}\label{eq:de}
de=Tds-Pd\mbox{\it v}+\mu_id\nu_i.
\end{equation}
However, we cannot write a similar expression for $E^{\rm cryst}$, because of its explicit dependence on $N_L$. 
Using the fact that $N_L ds$=$dS^{\rm cryst}-sdN_L$, $N_L d\mbox{\it v}$=$dV^{\rm cryst}-\mbox{\it v}dN_L$ and $N_L d\nu_i$=$dN_i^{\rm cryst}-\nu_idN_L$, 
we can rewrite Eqn.~\ref{eq:dE} as
\bey{eq:varE}
dE^{\rm cryst} &=& \left(TdS^{\rm cryst}-P^{\rm cryst}dV{\rm cryst}+\mu_idN_i^{\rm cryst}\right)\nonumber\\
&+&\left(e-Ts-\mu_i\nu_i +P^{\rm cryst}\mbox{\it v}\right)dN_L ,
\eey
which follows from the explicit expression for $de$ (Eqn.~\ref{eq:de}), and the expressions for  $dS^{\rm cryst}$, $dV^{\rm cryst}$ and $dN_i^{\rm cryst}$ above Eqn.~\ref{eq:varE}.
Note that $e-Ts-\mu_i\nu_i=\Omega^{\rm cryst}/N_L$. 
We can then write the variation of the grand potential of the crystal as:
\bey{eq:dOmega}
d\Omega^{\rm cryst} &=& dE^{\rm cryst} -TdS^{\rm cryst} +\mu_idN_i^{\rm cryst} \\
&=& -P^{\rm cryst} dV^{\rm cryst}+(\Omega^{\rm cryst}/N_L+P^{\rm cryst}\mbox{\it v}) dN_L . \nonumber
\eey
The condition for (unstable) equilibrium is then:
\bey{eq:dom_tot}
d\Omega^{\rm tot} &=& -P^{\rm cryst}dV^{\rm cryst} +
(\Omega^{\rm cryst}/N_L+P^{\rm cryst}\mbox{\it v}) dN_L \nonumber\\
 &-&P^{\rm liq}dV^{\rm liq} +d\Omega^{\rm surf}=0.
\eey
Our next step is to derive an expression for the radius of the critical nucleus. 
At this stage, the expression $dV^{\rm cryst}$= $N_L d\mbox{\it v}+ \mbox{\it v}dN_L$  (eqn.~\ref{eq:dV}) becomes important because it expresses the fact that there are two ways to change  size of a nucleus: either by changing the volume $\mbox{\it v}$ of the unit cell, while keeping $N_L$, the number of lattice points, fixed, or by changing $N_L$ at constant $\mbox{\it v}$.
It is convenient to consider these two scenarios separately. 
When a crystal nucleus grows, we are adding lattice sites, \ie we are changing $N_L$, and when a crystal grows/shrinks at constant number of lattice sites, we are changing $\mbox{\it v}$.

We will consider these two scenarios separately.
First I consider the case that $v$ is fixed, and $N_L$ can vary.
\subsubsection{Fixed unit-cell volume}
In that case, 
\bey{eq:DOM_barrier_1}
d\Omega^{\rm tot}&=&\omega^{\rm cryst}dV^{\rm cryst}+\omega^{\rm liq}dV^{\rm liq}+\dfdx{\Omega^{\rm surf}}{V^{\rm cryst}}dV^{\rm cryst}\nonumber \\
&=& \left(\omega^{\rm cryst}-\omega^{\rm liq}+\dfdx{A\omega^{\rm surf}}{V^{\rm cryst}}\right)dV^{\rm cryst},
\eey
where I have used the fact that $dV^{\rm liq}=-dV^{\rm cryst}$ and that,  at constant $\mu,T,\mbox{\it v}$, $\omega^{\rm cryst}$ and $\omega^{\rm liq}$ are constant, and $A\omega^{\rm surf}$ depends only on the radius $r$ of the nucleus.
As we have kept $\mbox{\it v}$ fixed, $dV^{\rm cryst}= \mbox{\it v}dN_L$, and hence Eqn.~\ref{eq:DOM_barrier_1} can be written as
\beq{eq:mu_latt}
\dfdx{\Omega^{\rm tot}}{N_L}\equiv \mu_{\rm lat}=0,
\eeq
where we have {\em defined} $\mu_{\rm lat}$ as the ``chemical potential'' of the lattice sites. 
As the number of lattice sites is not conserved, $\mu{\rm lat}$ should be zero ... and it is.

We can now compute the free-energy cost $W$ of making a nucleus with radius $r$ by integrating Eqn.~\ref{eq:DOM_barrier_1} from $r=0$ to the radius  $r$.
\bey{eq:W1}
W &=& \left(\omega^{\rm cryst}-\omega^{\rm liq}\right)V^{\rm cryst}+A\omega^{\rm surf}\\
&=& \left(\omega^{\rm cryst}+P^{\rm liq}\right)V^{\rm cryst}+A\omega^{\rm surf} .\nonumber
\eey
We note that the choice of the radius dividing surface is still unspecified.
To proceed, I note that, at the top of the nucleation barrier, we must have
\bey{eq:top}
\dfdx{W}{V^{\rm cryst}} &=& \left(\omega^{\rm cryst}- \omega^{\rm liq}\right)\\
&+&(2/r)\left(\omega^{\rm surf}+(r/2)\dfdx{\omega^{\rm surf}}{r}\right)=0 .\nonumber
\eey
Until now, we have not specified how we define the radius of the critical nucleus, $r^*$.
A convenient (but not unique) choice is to  impose the condition
\beq{eq:dodr_zero}
\dfdx{\omega^{\rm surf}}{r}_{r^*}=0,
\eeq
where it should be stressed that the derivative is take at constant $\mbox{\it v}$. 
Then 
\beq{eq:top2}
\dfdx{W}{V^{\rm cryst}} = \left(\omega^{\rm cryst}-\omega^{\rm liq}\right)+\frac{2\omega^{\rm surf}}{r}=0 ,
\eeq
and hence
\begin{equation}
\omega^{\rm surf}= (r/2)(\omega^{\rm liq}-\omega^{\rm cryst}) .
\end{equation}
Inserting $\omega^{\rm surf}$ in Eqn.~\ref{eq:W1}, we get
\begin{equation}
r^* = \left(\frac{(3W/2\pi)}{\omega^{\rm liq}-\omega^{\rm cryst}}\right)^{1/3}
\end{equation}
Defining the area of the critical nucleus as $A^*$, and using the familiar notation\; $\omega^{\rm surf}\equiv \sigma$  we then get
\begin{equation}\label{eq:Mullins-barrier}
W=\frac{\sigma A^*}{3} ,
\end{equation}
which is Gibbs's famous result.

Note that Eqn.~\ref{eq:Mullins-barrier} for the nucleation barrier looks  the same as Eqn.~\ref{eq:barrier-2}.
However, the meaning is different, because in one case, $A$ and $\sigma$ refer to the surface of tension, and in the other to the equimolar surface. 
Of course, the barrier height itself does not depend on our choice of dividing  surface.
To reconcile the two expressions, we must account for curvature corrections to the surface free energy.
If we take the curvature dependence of the surface free energy into account in Eqn.~\ref{eq:barrier-2}, we get (approximately):
\begin{eqnarray}\label{eq:barrier-2a}
\Delta F_{\rm critical} &=&  \left(\frac{2\sigma _\infty}{r^e_c} +\frac{C_1}{(r^e_c)^2}\right)V^e_c  \nonumber\\
&+& \int_0^{V^e_c}  \left(\frac{2\sigma _\infty}{r}+\frac{C_1}{r^2}\right)  \dd V_{\rm solid}+4\pi C_2\nonumber\\
&=& \frac{\sigma _\infty A^e_c}{3}\left(1+  \frac{2C_1}{r^e_c\sigma _\infty}\right) +4\pi C_2\;,
\end{eqnarray}
where $\sigma _\infty$ denotes the free-energy density of a flat surface, the superscript $e$ refers to the equimolar surface, and the subscript $c$ refers to the critical nucleus.
The curvature-related coefficients $C_1$ and $C_2$ have been defined in Eqn.~\ref{eq:curved-surface}.
Of course, the height of the nucleation barrier does not depend on the choice of the dividing surface.
We can also express eqn.~\ref{eq:barrier-2a} also in terms of the radius of the surface of tension, using 
$\delta\equiv r_{\rm equimolar}-r_{\rm tension}$.
Note that in this ``macroscopic'' (thermodynamic) description,   the term with $C_2$ appears because, even though it does not influence the variation of $F_s$ with $V$, it is zero in the absence of a nucleus, and non-zero as soon as a nucleus appears (see refs.~\onlinecite{tenwolde1998,mcgraw1997}).

Many modern texts on homogeneous nucleation employ expressions for the nucleation barrier involving the chemical potential difference between solid and fluid.
As should be clear from the discussion in the previous sections, it is dangerous to make use of this chemical potential difference in the context of crystal nucleation.

\subsubsection{Fixed number of lattice sites}\label{ssec:fixed}
At fixed $N_L$ and variable $v$, we start from Eqn.~\ref{eq:dom_tot}, written as:

\bey{eq:dom_constv}
d\Omega^{\rm tot} &=& \left(-P^{\rm cryst}+  P^{\rm liq}\right)dV^{\rm cryst} 
 +d\Omega^{\rm surf}=0 ,
\eey
where 
\begin{equation}
dV^{\rm cryst} =N_L d\mbox{\it v}.
\end{equation}

This is the volume change is due to expansion or compression of the lattice, whilst the volume change at constant $v$ is due to accretion/dissolution. 
From Eqn.~\ref{eq:dom_constv} it follows that
\beq{eq:laplace}
\left(P^{\rm cryst}-P^{\rm liq}\right)=\frac{2}{r}\left(\sigma+(3\mbox{\it v}/2)\dfdx{\sigma}{\mbox{\it v}}\right).
\eeq
As before, I define the surface tension $t$ as
\begin{equation}
t\equiv \sigma+(3\mbox{\it v}/2)\dfdx{\sigma}{\mbox{\it v}}.
\end{equation}
Note that the surface tension includes a term $d\sigma/d \mbox{\it v}$: this term did not appear in Eqn.~\ref{eq:dodr_zero}, because there I considered the variation of $\sigma$ with $r$ {\em at constant $\mbox{\it v}$}.
The result is that for crystals, the surface tension $t$ is not equal to the surface free-energy density $\sigma$, and the difference may be large~\cite{cacciuto2005stresses,montero2020young}.
Eqn.~\ref{eq:laplace} shows that the "Laplace" pressure of a crystal inside a fluid contains a term that depends on the stretching/compression of the surface, which is why the pressure inside a crystal nucleus is {\em not} simply $2\sigma/r$.

It is interesting to compare the above results with those for fluid droplets. It is usually argued (as I did above) that  $\sigma=t$ for planar fluid-fluid interfaces. 
But for small droplets, life is not that simple:   Buff~\cite{buff1955} has given an explicit statistical-mechanical expression for the surface tension of spherical droplets containing particles that interact through pair potentials.
Buff defined the location of  the ``mechanical surface of stress'' in terms of an integral over the tangential part of the (Irving-Kirkwood) stress tensor, arriving at an expression that is analogous to Gibbs definition of the equimolar dividing surface (but not quite as, unlike the density, the stress tensor is not uniquely defined).
Buff's expression for the ``mechanical'' surface tension  of fluid-fluid interfaces has been used by ten Wolde \etal~\cite{tenwolde1998}, who found that this ``mechanical surface of stress'' did not coincide with the thermodynamic surface of tension -- not even for a fluid!
It is for this reason that I wrote in the beginning of this article that the correct description of the mechano-chemical equilibrium between a crystalline solid and a fluid  only {\em seemed} to require much more care than the description of equilibrium between two coexisting fluid phases. 

For crystals, the same stress-tensor integration should yield $t$, not $\sigma$.
If, in the spirit of Gibbs, we could assume that the surface stress is localized in a mathematical spherical surfaces (the  ``surface of  mechanical stress''), which need {\em not} be the same as the thermodynamic surface of tension, then the Laplace pressure inside a crystal should be given by 
 \begin{equation}
 \left(P^{\rm cryst}-P^{\rm liq}\right)\equiv \frac{2t}{r_{\rm stress}},
 \end{equation}
which defines $r_{\rm stress}$,  the radius of the surface of mechanical stress.
One problem with this definition is that, unlike particle density, the microscopic stress tensor is not uniquely defined. 
 This ambiguity has no consequences for the prediction of the surface tension of a planar surface~\cite{schofield1982}, but for curved surfaces different choices for the form of the stress tensor result in different predictions of $t$, and therefore for $r_{\rm stress}$, the radius of this surface of stress.
Moreover,  $r_{\rm stress}$ is in general different from both the radius of the equimolar dividing surface, and that of the (thermodynamic) surface of  tension.
The fact that ref.~\onlinecite{tenwolde1998} found that the surface of tension and the surface of stress do not coincide for small fluid droplets supports an earlier theoretical prediction by Blokhuis and Bedeaux~\cite{blokhuis1992} that, even for fluid droplets, we should distinguish between surface stress and surface tension.

This observation may not be all that surprising because, once droplets are no longer large compared with the range of the oscillations in the radial distribution function, the surface of a fluid droplet is to some extent structured, and expanding/compressing the surface may change this structure. 

As $r_{\rm stress}$ is not uniquely defined, one might as well choose it equal to the radius of the surface of (thermodynamic) tension. 
Only in that case is it meaningful to compare the surface free-energy density $\sigma$ and the surface stress $t$. 
Implicitly, this approach was followed in the thermodynamic treatment of $\sigma$ and $t$.

\subsection{Vacancies and the chemical potential inside a nucleus}\label{ssec:chempot}
The advantage of the Mullins approach is that it follows the Grand-Canonical route and hence, by construction, chemical equilibrium of every species is imposed  throughout the system.
However, and this has been the source of much confusion~\footnote{Actually, not all that much, because the problem is not widely recognized.}, the chemical potential of a bulk solid at the temperature and pressure prevailing inside the nucleus, is {\em not} equal to the chemical potential of the fluid phase (see Eqn.~\ref{eq:mus-ne-mul}). 
So, how can this be?

Before going into more detail, I should point out that in finite systems, the thermodynamic relation between the chemical potential and the pressure is no longer the standard Gibbs-Duhem equation $Nd\mu = VdP$ (for a one-component system).
Rather, as Hill has shown~\cite{hill1994thermodynamics,bedeaux2023}, for small systems there is an additional term in the ``Hill-Gibbs-Duhem'' equation due to the ``sub-division energy''. 
So, at the level of Hill's thermodynamics,  there is no problem -- but it would be useful to gain a more detailed  understanding how, in small, but not even very small~\cite{montero2020young} crystals,  the chemical potential can ``uncouple'' from the pressure. 

A crucial observation is that, in a statistical mechanical description of crystals at finite temperature, the existence of point defects (vacancies and interstitials),  is unavoidable.
To see this, consider a solid with $M$ lattice sites at constant chemical potential.
I focus on a one-component solid, because that is good enough for the argument. 
We can write the grand partition function of this crystal:
\bey{eq:GC_crystal}
\Xi_{\rm solid} &=&\sum_{N=0}^\infty Q_M(N,V,T)e^{\beta\mu N}\nonumber\\
&\approx&  \sum_{N=0}^M Q_M(N,V,T)e^{\beta\mu N},
\eey
where $Q_M(N,V,T)$ is the canonical partition function of a crystal with $N$ particles distributed over $M$ lattice sites.
In the second line of Eqn.~\ref{eq:GC_crystal}, I have ignored interstitials and multiple occupancy of lattice cells, which is almost always justified (but not essential for the argument).
If we define the number of vacancies as $n\equiv M-N$, we can write
\bey{eq:GC_crystalb}
\Xi_{\rm solid} =e^{\beta\mu M}\sum_{n=0}^M Q_M(n,V,T)e^{-\beta\mu n}
\eey
In what follows, I will assume (again for simplicity) that the vacancy concentration is very low, and the interaction between vacancies can be ignored.
This assumption may not always be correct, but the more general case of interacting vacancies, while straightforward, is less transparent. 
Now, if we express $Q_M(n,V,T)$ in terms of the free energy to create $n$ non-interacting vacancies:
\beq{eq:vacancies}
\frac{Q_M(n,V,T)}{Q_M(0,V,T)}\equiv  \binom{M}{n}e^{-\beta nf_v(V,T)}  \;,
\eeq
where $f_v(V,T)$ denotes the free-energy cost to create a single vacancy. 
Now we can write Eqn.~\ref{eq:GC_crystalb} as
\bey{eq:GC_2}
\Xi_{\rm solid} &=&Q_M(0,V,T)e^{\beta\mu M}\sum_{n=0}^M \binom{M}{n}e^{-\beta n(f_v(V,T)+\mu)}\nonumber\\
&=& Q_M(0,V,T)e^{\beta\mu M}\left(1+e^{-\beta(f_v(V,T)+\mu)}\right)^M.
\eey
Note that  the free-energy cost of creating a single vacancy at a specific lattice site in the crystal is equal and opposite to the free-energy cost of adding a particle if there were a vacancy at that location~\cite{pronk2001,purohitKofke2018}).  
In a perfect, defect-free crystal, {\em there are no vacancies where particles can be inserted}.
Establishing  chemical equilibrium between a reservoir and a crystal requires a non-vanishing concentration of vacancies.
But once vacancies are possible, there is no problem with the chemical potential of the particles inside the crystal: it is simply equal to the chemical potential of the reservoir.
It is important to notice that it is immaterial that the Laplace pressure is equal to $2t/r$ rather than $2\sigma/r$: the only thing that matters is that the crystal is in equilibrium with the reservoir.
This is a somewhat counterintuitive observation. 
To clarify it, consider a thought experiment where we have a crystallite of surfactant molecules on an air-water interface (the ```thought'' experiment could also be a simulation~\cite{ciccotti1979}). 
In equilibrium, this quasi-2D crystallite will contain some vacancies.
We also consider that there is a finite equilibrium concentration of the surfactants in the water and/or the air -- but let us just consider free surfactants in solution.
The presence of vacancies in the surfactant crystals makes it possible to establish equilibrium between the dissolved surfactants and those in the crystal phase, no matter what the equilibrium vacancy concentration is, as long as it is non-zero. 
Normally, this equilibration will be slow as it required diffusion in the crystalline solid, but in our 2D surfactant crystal, equilibration can be reached through the third dimension, without passing through the circumference of the crystal that traces the crystal-fluid interface, although that route of equilibration is also possible.

Changing the Laplace pressure in the 2D crystal will change the equilibrium concentration of vacancies,  but {\em not} the equality of chemical potential between crystallite and free surfactants.
Addition of particles at the boundary of the crystal is different: when  particles are added on  surface of a crystal, they do so by increasing the number of lattice sites.
In equilibrium, the rate of particle addition and removal balance.
This means that the chemical potential of particles in the surface layer mut have the same chemical potential as those in the fluid.
Due to the vacancy-mediated mobility of particles in the crystal,  the chemical potential inside of a crystallite will (eventually) equilibrate with the surface, and therefor with the fluid. 

This observation is probably what Gibbs meant when he wrote:
{\em "...the temperature and pressure and the potentials for all the actual components of the solid must have a constant value in the solid  at the surface where it meets the fluid}.  
In fact, when describing multi-component systems, for which there is a net surface excess for all but one of the components, Gibbs explicitly states that  the excess components in the dividing surface have the same chemical potential as in the bulk fluid (ref.~\onlinecite{gibbs1928}, Eqn. 508).
But  that chemical potential is not the same as the reversible work needed to add a particle from the fluid to  a {\em bulk} crystal at the temperature and pressure that prevail inside the nucleus. 

In experiments, the vacancy concentration can differ -- unintentionally or intentionally -- from its equilibrium value.
This has consequences for the chemical potential of the molecules in the solid.
To see this, consider the variation of the Gibbs free energy of a bulk crystal with $N_L$ lattice sites as:
\begin{equation}
dG =-SdT +VdP +\mu_i dN_i + \mu_L dN_L ,
\end{equation}
where, as before, we have defined $\mu_L$ as the ``chemical potential'' of the lattice sites.
Of course, in a defect-free lattice, $N_L$ is linearly related to $N_i$, but if the occupancy of lattice sites can vary, $N_L$ is an independent thermodynamic variable.
In equilibrium, lattice sites can be created or deleted at no free-energy cost, and hence $\mu_L$=0.
But in a metastable crystal with a non-equilibrium value of $N_L$, the corresponding value of $\mu_L\ne 0$. 
In a simulation, we can  determine the value of $N_i/N_L$ for which $\mu_L$=0 (see \eg~\onlinecite{mladek2007}). 
Even if $\mu_L\ne 0$, \ie there are too few or too many vacancies,  we can still  compute $\mu_i$ by Widom's particle-insertion method, provided the vacancy concentration is non-zero.
However, the  value of $\mu$ thus ``measured'' will deviate from the equilibrium value.
Of course, in a Grand-Canonical simulation, the average concentration of vacancies, and thereby $\mu$ will regress to its equilibrium value.


\subsection{Measuring the effect of surface stress}
Gibbs assumed that, for real materials, one could ignore the effect of the difference between surface free-energy density ($\sigma $) and surface stress ($t$). 
There are, however, cases where the difference is important.  
In 2020, Montero de Hijes and  Vega  reported careful simulations that showed that the Laplace pressure inside a critical nucleus of a hard-sphere crystal is indeed negative~\cite{montero2020young}, whereas the surface free-energy density is necessarily positive.
An earlier, but less quantitative observation  by Cacciuto and Frenkel for a system of hard colloids with short-ranged attraction~\cite{cacciuto2005stresses} had also found that the Laplace pressure in small crystallites of this system could be negative, so much so that the density of small nuclei was {\em less} than that of larger nuclei.
So, what was unobservable in Gibbs's time, is now observable in simulations. 
\section{Possible experiments?}
It would seem difficult to measure in real experiments the effect of surface stress  on the properties of a solid inside a small crystallite in contact with a fluid, because small crystallites are always unstable: if they are smaller than the critical nucleus they dissolve, and if they are larger they grow to form a bulk crystal -
critical nuclei are in unstable equilibrium, meaning that they only survive until a fluctuation pushes them to grow or disappear. 

Surprisingly, it is possible to prepare {\em stable} critical nuclei.
One strategy is useful for simulations, but less so for experiment: it involves preparing a nucleus in a finite volume of the parent phase (see \eg \onlinecite{binder_nucleation}). 
When the nucleus grows, it consumes material in solution until the concentration of the solution has decreased to the point where growing a larger nucleus no longer lowers the total free energy of the system.
Clearly, the size of the resulting stable nucleus depends on the initial volume of the parent phase. 
This approach is feasible in simulations, but much less so in experiments on microscopic nuclei (may be in microfluidic droplets). 

There is however another possibility: one can prepare a crystallite in a cylindrical pore or in a slit. 
In supersaturated solutions, the crystallite in the pore grows until its radius of curvature has decreased (!) to the value of the critical radius at that supersaturation of the solution~\cite{xiao2017}.
Then the fluid-exposed part of the nucleus behaves just like a critical nucleus.
For a crystallite growing out of a pore, Eqn.~\ref{eq:VvsA}, which relates the excess volume $\Delta V$ to the excess surface area $\Delta A$ of the nucleus with respect to the original flat-topped cylindrical crystallite still holds. 
Hence, the entire analysis relating the surface stress to the Laplace pressure still applies (for more details, see ref.~\onlinecite{xiao2017}).
Therefore, growing a crystallite that "wets" a cylindrical pore, but not the top surface of the flat substrate, makes it possible to create part of a stable critical nucleus. 
One can then measure the lattice spacing inside such a nucleus and deduce the Laplace pressure. 
I am unaware of publications reporting such experiments. 
\section{Conclusions}
This paper has a few takeaway messages. 
\begin{itemize}
    \item The first message, although ``obvious'', is highly non-trivial: \textit{yes, the chemical potential inside a critical nucleus is equal to that of the fluid parent phase with which it is in unstable equilibrium, meaning that the two are in ``osmotic'' equilibrium, \ie the condition where there is no spontaneous mass flux between fluid and crystallite}.
    \item This ``obvious'' result is non-trivial because if a macroscopic (bulk) crystal  were kept at the  pressure $P_{\rm bulk\; coex}+2t/r$ that prevails well-inside a critical nucleus, it  would {\em not} have the same chemical potential as the particles in the critical nucleus.
    The reason is precisely that chemical potential inside a crystal is set by the chemical potential at its surface, where (as Gibbs pointed out) it is equal to that of the adjacent fluid, which can be viewed as a grand-canonical reservoir at constant chemical potential.
    At a pressure $P_{\rm bulk\;coex}+2t/r$, the chemical potential in the fluid would not result in coexistence with the critical nucleus, because that requires a fluid pressure $P_{\rm bulk\;coex}+2\sigma/r$. 
    I am well-aware that this result is counter-intuitive, because, the lattice spacing of the bulk crystal and the core of the crystal nucleus are the same at a pressure $P_{\rm bulk\;coex}+2t/r$.

    \item If the fluid-solid surface free-energy density $\sigma$ would be equal to the surface tension $t$, the chemical-potential paradox would go away.  
    But for crystals, the surface free-energy density is, in general, not equal to the surface tension.
    In contrast, for fluids,  $\sigma$ and $t$ are expected to be the same. 
    But this seems no longer true for droplets with a radius that is not much larger than the size of the constituent particles~\cite{tenwolde1998}.
    \item
    Only due the presence of a (small) equilibrium concentration of vacancies can the equilibrium between a crystal nucleus and fluid be established. 
    \item hopefully, the conclusions of this paper can be tested in experiment -- but it will not be easy. 
\end{itemize}
\nocite{mesarovic2016,sprik2025}
\section*{Acknowledgments}
This article is dedicated to Christoph Dellago. 
In 1997, I wrote a report supporting his applications for an Erwin Schr\"{o}dinger fellowship to work with David Chandler. 
I have no doubt that Christoph would have been a very successful scientist even if I had not written this report. 
But  it still makes me feel happy that I was "on the right side of history". 
Of course, times have changed, as history seems no longer to be on the side of science. 
But science is always on the side of evidence and, in the long run (if there is a long run), evidence should win. 
It is wonderful to work in the same academic universe as Christoph Dellago: for me, he is a role model.

I discussed the topic of this paper extensively with my part-time room mate Michiel Sprik. 
His stern but fair comments convinced me to delete about half of the original text and rewrite the other half. 
I also discussed part of the work with Dick Bedeaux, Signe Kjelstrup,  David Reguera and, in particular, Aleks Reinhardt, David Kofke and Pieter Rein ten Wolde:  I gratefully acknowledge all comments and suggestions but  of course all remaining mistakes are mine, and mine alone.
\clearpage
\bibliographystyle{unsrt}
\bibliography{gibbs_lib.bib}
\end{document}